\documentclass[12pt,preprint]{revtex4}
\usepackage{graphicx} 
\usepackage{amsmath}
\usepackage{amsfonts,amssymb}
\usepackage[matrix,arrow,curve,frame,poly,arc]{xy}

\def\be{\begin{equation}}
\def\ee{\end{equation}}
\def\ba{\begin{eqnarray}}
\def\ea{\end{eqnarray}}

\begin{document}
\bibliographystyle{plainnat}

\title{The system of two parabolic equations as an integrable system. }

\date{30.07.2017}

\author{Maria Shubina }

\email{yurova-m@rambler.ru}

\affiliation{Skobeltsyn Institute of Nuclear Physics\\Lomonosov Moscow State University
\\ Leninskie gory, GSP-1, Moscow 119991, Russian Federation}


\begin{abstract}

In this paper we present a system of two nonlinear partial differential equations of the second order, depending on the time and one spatial coordinate. It can be written as a system of two Burgers equations, which allows one to immediately conclude that it is integrable. This system seems to be new, and its consideration can be of independent interest.

\end{abstract}

\keywords{integrable system, parabolic-parabolic system, exact solution, Patlak-Keller-Segel model}

\maketitle

\section{Introduction}
\label{intro}

In this paper we consider a system of two parabolic nonlinear differential equations for functions that depend on time and one spatial coordinate:
\be \left\{
\begin{aligned}
u_{t} & =  u_{xx} - \eta \, (u \, \phi_{x})_{x}  \\ 
\phi_{t} & =  \phi_{xx} - \frac{\eta }{2} \,\left(  (\phi_{x})^{2} + u^{2} \right), 
\end{aligned}
\right.
\ee
where $ \eta > 0 $, $ x \in \Re, t \geq 0 $, $ u=u(x,t) $, $ \phi = \phi(x,t) $. 

The aim of this paper is to consider this system, which can be represented as a system of two Burgers equations 
\be
g_{t}  =  g_{xx} - \eta \, g \, g_{x}. 
\ee
Consequently it can be linearized and it is integrable \cite{Conte&Musette}, and its general solution, corresponding to arbitrary initial conditions and regular at infinity, is known. The solution in terms of traveling wave variable will be presented below. This system seems to be new, see Shabat \cite{Shabat}. 

\section{Chemotaxis and integrability}
\label{sec:1}

System (1) appeared as a result of the desire to modify the system of chemotaxis, so as to obtain an integrable model. Chemotaxis plays an important role in many biological and medical fields such as embryogenesis, immunology, cancer growth. The macroscopic classical model of chemotaxis was proposed by Patlak in 1953 \cite{P} and by Keller and Segel in the 1970s \cite{KS1}-\cite{KS2}. This model describes the space-time evolution of a cells density $ u(t,\overrightarrow{r}) $ and a concentration of a chemical substance $ v(t,\overrightarrow{r}) $. The most frequently considered 1D models have the form:
\be \left\{
\begin{aligned}
u_{t} & =  u_{xx} - \eta \, (u \, \phi(v)_{x})_{x}  \\ 
v_{t} & = \alpha v_{xx} + f(u, v), 
\end{aligned}
\right.
\ee
where  $ \eta, \alpha > 0 $, $ u=u(x,t) $, $ v=v(x,t) $; the function $ \phi (v) $ is the chemosensitivity function and $ f(u, v) $ characterizes the chemical growth and degradation. These functions are taken in different forms, for example $ \phi (v) \sim v $, $ \phi (v) \sim  \ln v $, $ f(u, v)= \sigma u - \beta v  $ or $f(u, v) = - v^{m} u -  \beta v $, $ \sigma $, $ \beta $, $ m $ are constants \cite{Hillen&Painter}, \cite{Wang}, and \cite{H1}-- \cite{TF}. One can show \cite{Shubina_arxiv}, that for any of these functions system (3) does not pass the Painlev\'e test \cite{Conte&Musette}, \cite{WTC}--\cite{Kudryashov}, and, therefore, is not integrable. However, the reduction to the traveling wave variable allows to obtain exact solutions in closed form, and the reductions themselves pass the Painlev\'e  test for ordinary differential equations for certain values ​​of parameters.

Everything described led us to the idea of ​​modifying system (3) in some way, in order to obtain an integrable model. Of course, this work requires further development, but it allowed us to get, as it seems to us, a new integrable system.

\section{System (1) and the Burgers equation}
\label{sec:2}

Writing the system of two Burgers equations (2) for the functions
\be
g_{\pm} = u \pm \phi_{x},
\ee
we obtain system (1). Further, if we want to "approach" system (3) we pass from $ \phi $ to $ v $, where $ \phi = \phi(v) $, and for $ \phi_{v} \neq 0 $ we obtain the following system:
\be \left\{
\begin{aligned}
u_{t} & =  u_{xx} - \eta \, (u \, \phi(v)_{x})_{x}  \\ 
v_{t} & =  v_{xx} + \left(  \frac{\phi_{vv}}{\phi_{v}}  - \frac{\eta }{2} \, \phi_{v}  \right)\, (v_{x})^{2}  - \frac{\eta }{2 \phi_{v}} \, u^{2}.  
\end{aligned}
\right.
\ee
Let $ u $ and $ v $ denote the same thing as in system (3). As is easy to see, here the first equation containing the nonlinear term (chemotaxis) coincides with the first equation in (3), but the second equation in (5) already differs significantly from the second in (3). If we are interested in the case $ \eta > 0 $ (the so-called "positive" chemotaxis \cite {Ni}), then the bracket before $ (v_ {x}) ^ {2} $ is not zero. Next, we write down the solution of system (5) for the traveling wave variable and, analyzing the asymptotics of the functions $ u $ and $ v $, we define the suitable form of $ \phi (v)$.

\section{Exact solution}
\label{sec:3}

If in (5) the functions $ u $ and $ v $ describe the quantities measured in the experiment, then we are interested primarily in non-negative solutions that are bounded on the entire domain of variation of coordinates and time. We find the solution in terms of the traveling wave variable $ y = x - c t $. For the function $ v $, the solutions will be regular if we put $ \phi (v) = \ln v $. Then we obtain:
\ba
u (y) & = & \frac{A(C_{-} - C_{+})}{4 \sqrt{C_{-}C_{+}}} \,\, sech\left( \frac{A \eta}{2}y + \frac{1}{2} \ln C_{-} \right)\, sech\left( \frac{A \eta}{2}y + \frac{1}{2} \ln C_{+} \right)    \\ 
v (y) & = & \left[ \frac{1}{4 \sqrt{C_{-}C_{+}}} \, e^{cy} \, sech\left( \frac{A \eta}{2}y + \frac{1}{2} \ln C_{-} \right)\, sech\left( \frac{A \eta}{2}y + \frac{1}{2} \ln C_{+} \right) \right] ^{\frac{1}{\eta}}  
\ea
where $ A $, $ C_{\pm} $ are positive constants, satisfying the following conditions $ A \geq \dfrac{|c|}{\eta} $ and $ C_{-} > C_{+} $. For $ A < \dfrac{|c|}{\eta} $ the solution diverges at $ cy \rightarrow \infty $. 
It should be noted that for $ \phi(v) = v $, that is, when system (5) becomes identical to (1), the solution for $ v $ diverges.

\section{Conclusion}
\label{sec:5}

We obtained an integrable system that reduces to the system of Burgers equations, and also received bounded positive solutions in terms of the traveling wave variable. The work involves further development, in particular, a more detailed analysis of the possible interpretation of functions whose evolution is described by the system under consideration.

\end{document}